\def\be{\begin{equation}}
\def\ee{\end{equation}}
\def\dd{{\rm d}}
\def\etal{{\rm et al.\thinspace}}
\def\eg{{\rm e.g.\ }}

\def\ie{{\rm i.e.\ }}
\def\Fig{Figure}

\def\Eq{Equation}
\def\Eqs{Equations}
\def\Sec{Section}

\def\spose#1{\hbox to 0pt{#1\hss}}
\def\approxlt{\mathrel{\spose{\lower 3pt\hbox{$\sim$}}
	\raise 2.0pt\hbox{$<$}}}
\def\approxgt{\mathrel{\spose{\lower 3pt\hbox{$\sim$}}
	\raise 2.0pt\hbox{$>$}}}
\def\approxpropto{\mathrel{\spose{\lower 3pt\hbox{$\sim$}}
	\raise 2.0pt\hbox{$\propto$}}}
\def\bfit{\beta_{\rm fit}}
\def\rvec{{\bf r}}
\def\uvec{{\bf u}}
\documentstyle[11pt,paspconf,psfig]{article}
\psdraftbox

\begin{document}

\title{Multiphase Cooling Flows}
\author{Peter A. Thomas}
\affil{Astronomy Centre, Dept.~of Physics and Astronomy, University of
Sussex, Falmer, Brighton, BN1\,9QH, UK}

\begin{abstract}
I discuss the multiphase nature of the intracluster medium whose
neglect can lead to overestimates of the baryon fraction of clusters
by up to a factor of two.  The multiphase form of the cooling flow
equations are derived and reduced to a simple form for a wide class of
self-similar density distributions.  It is shown that steady-state
cooling flows are \emph{not} consistent with all possible emissivity
profiles which can therefore be used as a test of the theory.  In
combination, they provide strong constraints on the mass distribution
within the cooling radius.
\end{abstract}


\section{Introduction}
\label{sec:intro}

The multiphase nature of the intracluster medium (icm) in cooling
flows was demonstrated a decade ago when deprojections of X-ray
surface-brightness profiles showed that mass cools and is deposited
from the flow in a distributed manner, $\dot{M}\approxpropto r$ (\eg
Thomas, Fabian \& Nulsen 1987).  However, the complexity of the theory
and lack of data with high spatial resolution means that the
single-phase approximation is still widely adopted.  In this paper I
attempt to present the theory in a palatable form and give examples of
its application.

\section{The multiphase cooling flow equations}
\label{sec:multi}

The theory of multiphase cooling flows was set out by Nulsen (1986).
I rederive the equations in a slightly different form here in the hope
that they may prove more accessible.

\subsection{Derivation of the equations}

We assume an emulsion of density phases which comove with the flow.
The distribution is described by the volume fraction,
$f(\rho,\rvec,t)$, such that $f\,\dd\rho$ is the fractional volume
occupied by phases with densities in the range $\rho$ to
$\rho+\dd\rho$. Then $\int f\,\dd\rho=1$, and the mean density is
$\bar{\rho}=\int f\rho\,\dd\rho$.

Mass conservation gives
\be
{\partial\over\partial t}(\rho f)+\nabla.(\uvec\rho f)+
{\partial\over\partial\rho}(\dot\rho\rho f)=0,
\label{eq:mass}
\ee
where $\uvec$ is the rate of change of position and $\dot\rho$ is the
rate of change of density following the flow.  The final term in
\Eq~\ref{eq:mass} is the equivalent in density space of the
divergence in velocity space.

Integrating over all densities we obtain
\be
{\dot{\bar{\rho}}\over\bar{\rho}}+\nabla.\uvec+\beta=0,
\label{eq:mass1}
\ee
where
\be
\beta\equiv{1\over\bar{\rho}}\lim_{\rho\mapsto\infty}(\dot{\rho}\rho f).
\label{eq:beta}
\ee
This is equivalent to the usual single-phase equation (\eg Thomas
1988a) except that the mass deposition rate is specified in terms of
$f$ rather than being a free parameter.

To find how the volume fraction changes with time we use the energy
equation,
\be
{\dot{\rho}\over\rho}={1\over\gamma}{\dot{P}\over P}+
{\gamma-1\over\gamma}{n^2\Lambda\over P},
\label{eq:energy}
\ee
where $P$ is the pressure, $n^2\Lambda$ is the radiated power per unit
volume and $\gamma=5/3$ for an ionised plasma.

Over a wide temperature range appropriate to clusters the cooling
function can be approximated by a power-law, $\Lambda\propto
T^\alpha$, where $\alpha\approx0.5$.  Then
\Eq~\ref{eq:energy} can be simplified by moving to a new density
variable.  Writing
\be
\rho=\rho_0(\rvec,t)w^{-1/(2-\alpha)},
\label{eq:wdef}
\ee
we obtain
\be
\dot{w}=(2-\alpha)\left({\dot{\rho_0}\over\rho_0}-
{1\over\gamma}{\dot{P}\over P}\right)w-
(2-\alpha){\gamma-1\over\gamma}{n_0^2\Lambda(T_0)\over P}.
\label{eq:wdot}
\ee
If the adiabatic compression term is removed by setting
$P\propto\rho_0^\gamma$, then the energy equation takes a particularly
simple form, $\dot{w}=$constant.  However, a more useful
choice is to take $\rho_0\propto\bar{\rho}$.

From the final term of \Eq~\ref{eq:mass}, we see that at high density
when cooling is dominant, then $\dot{\rho}\rho f\sim\,$constant.  This
motivates the substitution
\be
f={(2-\alpha)\over\rho_0}w^{(4-\alpha)/(2-\alpha)}g(w,\rvec,t).
\ee
Then, using \Eqs~\ref{eq:mass}, \ref{eq:energy} and \ref{eq:wdot} we
obtain the following equation for the covariant derivate of $g$ (\ie
following the fluid flow):
\be
{\dot{g}\over g}+(3-\alpha){\dot{\rho_0}\over\rho_0}-
{2-\alpha\over\gamma}{\dot{P}\over P}+\nabla.\uvec=0.
\label{eq:gdot}
\ee

\subsection{The form of the density distribution}

In general $g$ is a complicated function of position and time.
However, we can look for solutions in which $g$ has a constant
functional form, $g=g(w)$.  Only the first term in \Eq~\ref{eq:gdot}
depends upon $w$.  Hence we require that $\dot{g}\equiv\dot{w}\,\dd
g/\dd w\propto g$.  There are two kinds of solution:
\begin{enumerate}
\item $g_\infty\propto\exp(-w)$.  This
is the most extended distribution which is convectively stable (it
gives $P\propto\rho_0^\gamma$).  It includes phases of arbitrarily low
density.
\item $g_k\propto(1-w)^{k-1}$, $0<w<1$; $k\geq1$.  
These solutions possess a minimum density, $\rho\geq\rho_0$.  $k=1$ is
the least extended, consisting solely of the power-law cooling tail.
As $k\mapsto\infty$ the solutions resemble $g_\infty$.
\end{enumerate}

For other forms of $g$ we must resort to numerical integration to
follow their evolution.  Thomas (1988b) looked at the steady-state
evolution of a range of distributions with a sharp cut-off at low
densities and reached the following conclusions:
\begin{itemize}
\item All distributions develop a high-density tail,
$f\sim\rho^{-(4-\alpha)}$, as they cool.
\item Sufficiently narrow distributions resemble the pure power-law
$g_1$ by the time they begin to be deposited.
\end{itemize}

$g_1$ and $g_\infty$ bound all reasonable solutions of the cooling
flow equations, be they self-similar in form or not.  I would also
argue that plausible formation histories for the icm make low values
of $k$ more likely than high values.

\section{The Baryon Catastrophe}
\label{sec:bc}

Modelling of the icm suggests that the baryon fraction in clusters is
of order 0.06$\,h^{-1.5}$ or more (\eg White \etal 1993, White \&
Fabian 1995), where $h=H_0/100\,$km\,s$^{-1}$Mpc$^{-1}$ is the
dimensionless Hubble parameter.  However, primordial nucleosynthesis
limits baryons to a small fraction of the critical density,
$\Omega_b<0.015\,h^{-2}$ (Walker \etal 1991).  If clusters contain a
representative mixture of material, then it would seem that the
Universe must be open with density parameter
$\Omega\approxlt0.25\,h^{-0.5}$.  This has been termed the Baryon
Catastrophe.

One flaw in the above argument is that the modelling is based on
single-phase models of the icm.  We know that there must be a wide
range of densities at the edge of the cooling flow and it is plausible
that the same is true throughout the cluster.  Moving to a multiphase
model can mitigate (but not eliminate) the Baryon Catastrophe.

First let me outline the determination of the gas and total masses of
a cluster in the single-phase case.  The equation of hydrostatic
support relates the mass of a spherical cluster to the properties of
its icm:
\be
{GM(<r)\over r}=-{1\over\rho}{\dd P\over\dd r},
\label{eq:hydrostatic}
\ee
where $G$ is the gravitational constant and $M$ is the total
mass within radius $r$.  To determine $M$ we need to know both the
density and pressure as a function of radius, whereas we usually
have only one constraint: the surface brightness profile can be
deprojected to give the emissivity, $\xi(r)=n^2\Lambda$.  To proceed
we assume a polytropic relation, $P\propto\rho^\Gamma$ where
$1\leq\Gamma\leq{5\over3}$ and normalise the average,
emission-weighted temperature, $T_X$, to some observed value.

When moving to a multiphase model the same procedure holds, except
that we have to average over all density phases.  Hence
$\rho\mapsto\bar\rho=\rho_0\int g\,\dd w$,
\be
\xi\mapsto\bar{n^2\Lambda}=n_0^2\Lambda(T_0)
\int w^{-(1-\alpha)/(2-\alpha)}g\,\dd w,
\ee
and so on.  The details of these calculations can be found in Gunn \&
Thomas (1996).  Suffice it to say that the form of the solutions is
unchanged but the gas mass is lowered and the total mass increased
over the single-phase case.  For $\alpha=0.5$ the baryon fractions are
lower by factors of 0.74 and 0.60 for the distributions $g_1$ and
$g_\infty$, respectively (these factors decrease slightly at lower
temperatures for which $\alpha$ is lower).

Finally, let me note that most models ignore complicating factors such
as magnetic fields and turbulence.  Both of these can act to support
the gas, thus raising the pressure in \Eq~\ref{eq:hydrostatic} and
lowering the gas fraction even further.  It is too early to conclude
that only low values of $\Omega$ are compatible with the data.

\section{Reconstruction of cluster mass profiles}
\label{sec:reconstruction}

The self-similar density distributions derived in \Sec~2 
lead to particularly simple forms of the steady-state cooling flow
equations.  I derive these below and then show how they can be
combined with the emissivity profile to provide strong constraints of
the mass distribution within the cooling radius.

\subsection{Theory}

Substituting the functional form of $g_k$ into 
\Eqs~\ref{eq:mass1}, \ref{eq:beta} and \ref{eq:gdot} we see that the
self-similar forms of the cooling flow equations are:
\be
{\dot{\rho_0}\over\rho_0}-{1\over\gamma}{\dot{P}\over P}-
{\beta\over(2-\alpha)k}=0
\ee
and
\be
{\dot{\rho_0}\over\rho_0}+\nabla.\uvec+\beta=0,
\ee
where
\be
\beta=(2-\alpha)k{\gamma-1\over\gamma}{n_0^2\Lambda(T_0)\over P}.
\ee
To these may be added the equation of hydrostatic support,
\be
\nabla\Phi+{\nabla P\over\bar\rho}=0,
\ee
where $\Phi$ is the gravitational potential.  I assume here that the inflow
is subsonic---this turns out to be a good approximation in all
multiphase cooling flow models.

In a steady-state and spherical symmetry the above equations reduce to
\be
{\dd\ln P\over\dd\ln r}=-2\Sigma,
\ee
\be
{\dd\ln\bar{\rho}\over\dd\ln r}=-{2\over\gamma}\Sigma-{\tau\over2-\alpha},
\ee
and
\be
{\dd\ln u\over\dd\ln r}=-2+{2\over\gamma}\Sigma+
\left({1\over2-\alpha}+k\right)\tau,
\ee
where
\be
\Sigma={GM\over2r}.{\mu m_H\over k_BT}
\ee
is the ratio of the virial to the thermal temperatures, and
\be
\tau=(2-\alpha){\gamma-1\over\gamma}{n_0^2\Lambda(T_0)\over P}
{r\over u}={1\over k}{\dd\ln\dot{M}\over\dd\ln r}
\ee
is the ratio of the inflow time to the constant-pressure cooling time.

Although there appear to be three equations here, the dimensionless
ratios $\Sigma$ and $\tau$ are the only important variables.  The
third equation merely acts as a scaling (for fixed $\tau$,
$\bar{\rho}^{2-\alpha}\propto p^{1-\alpha}u$).  Hence the physics can
be captured in just two equations:
\be
{\dd\ln\Sigma\over\dd\ln r}=\chi-1+2{\gamma-1\over\gamma}\Sigma-
{\tau\over2-\alpha}
\ee
and
\be
{\dd\ln\tau\over\dd\ln r}=3-
{2\over\gamma}\big[(3-\alpha)-\gamma(1-\alpha)\big]\Sigma-
\left({3-\alpha\over2-\alpha}+k\right)\tau
\ee
where $\chi\equiv\dd\ln M/\dd\ln r$.  The $g_\infty$
equations can be recovered by letting $k\mapsto\infty$ and using
$k\tau$ in place of $\tau$ as the second dimensionless variable.

The usual way of proceeding is to pick a functional form for the mass
profile, $\chi$, and then to solve for $\Sigma$ and $\tau$.  From this
one can generate and emissivity profile, $\xi(r)$, for comparison with
the data.  Alternatively, we can specify $\xi(r)$ and \emph{determine}
the mass profile.  Suppose that $\bfit\equiv-(1/6)\dd\ln\xi/\dd\ln r$
is known.  Then
\be
{\dd\ln\tau\over\dd\ln r}=3-
6{(3-\alpha)-\gamma(1-\alpha)\over2-\alpha+\alpha\gamma}\bfit-
\left({2\gamma\over(2-\alpha)(2-\alpha+\alpha\gamma)}+k\right)\tau.
\ee
Furthermore this is an eigenvalue problem: requiring that the solution
extend to $r=0$ fixes the outer boundary condition.  Hence we can
solve for $\tau$, $\Sigma$ and $\chi$.

We can get a good idea of the behaviour of the
solutions by looking at the case of constant $\bfit$.  Imposing the
physical constraints $\Sigma\geq 0$ (\ie a non-negative temperature)
and $\chi\geq0$ (\ie mass constant or increasing with radius)
restricts $\bfit$ to lie in the range
\be
{3\over2(5+3\,k)}\leq\bfit\leq{80+21\,k\over120+36\,k}
\ee
(in this expression and henceforth I set $\gamma=5/3$ and $\alpha=0.5$
rather than including them explicitly).  Thus steep emissivity
profiles, $\bfit\approxgt0.65$, are incompatible will all steady-state
cooling flow models (larger values can occur, however, outside the
cooling radius).  In addition, The inner value of $\bfit$ can be used
to constrain $k$ or, if we assume that the virial temperature drops
(\ie $\Sigma\mapsto0$) within the cluster core, to measure it.

Note that the solution for $\Sigma$ and $\tau$ does not depend upon
the normalisation of the gas and total gravitational masses.  If
desired, these can be determined by fixing the overall temperature and
luminosity.  The analysis of \Sec~\ref{sec:bc} shows that the gas
density will be slightly lower and the total mass density slightly
higher than in the equivalent single-phase analysis.

\subsection{Application to A85}

\begin{figure}
\psfig{figure=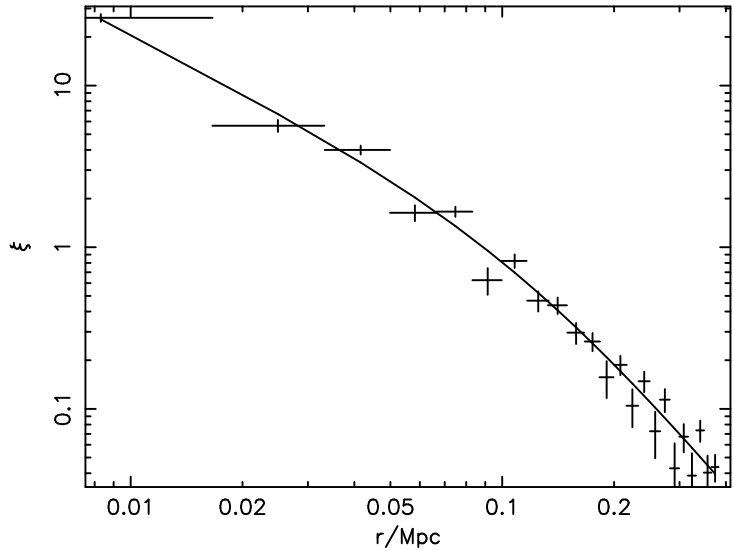,height=8cm,angle=270}
\caption{The emissivity profile for A85 in 12 arcsec bins.  The solid
line shows a broken power-law fit as described in the text.}
\label{fig:a85xi}
\end{figure}

\begin{figure}
\parbox{7.2truecm}{
\psfig{figure=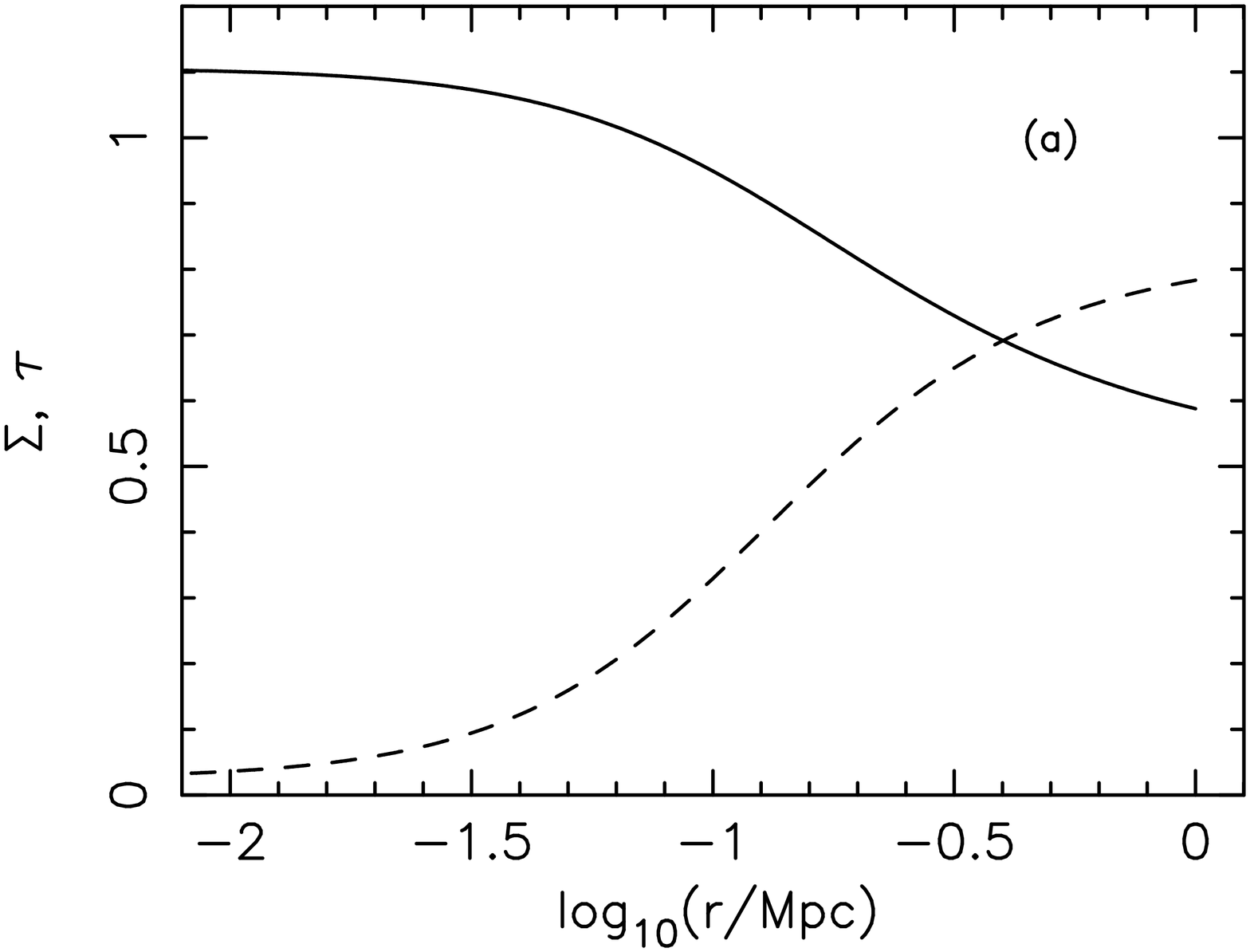,width=7.2cm,angle=270}
}\parbox{7.2truecm}{
\psfig{figure=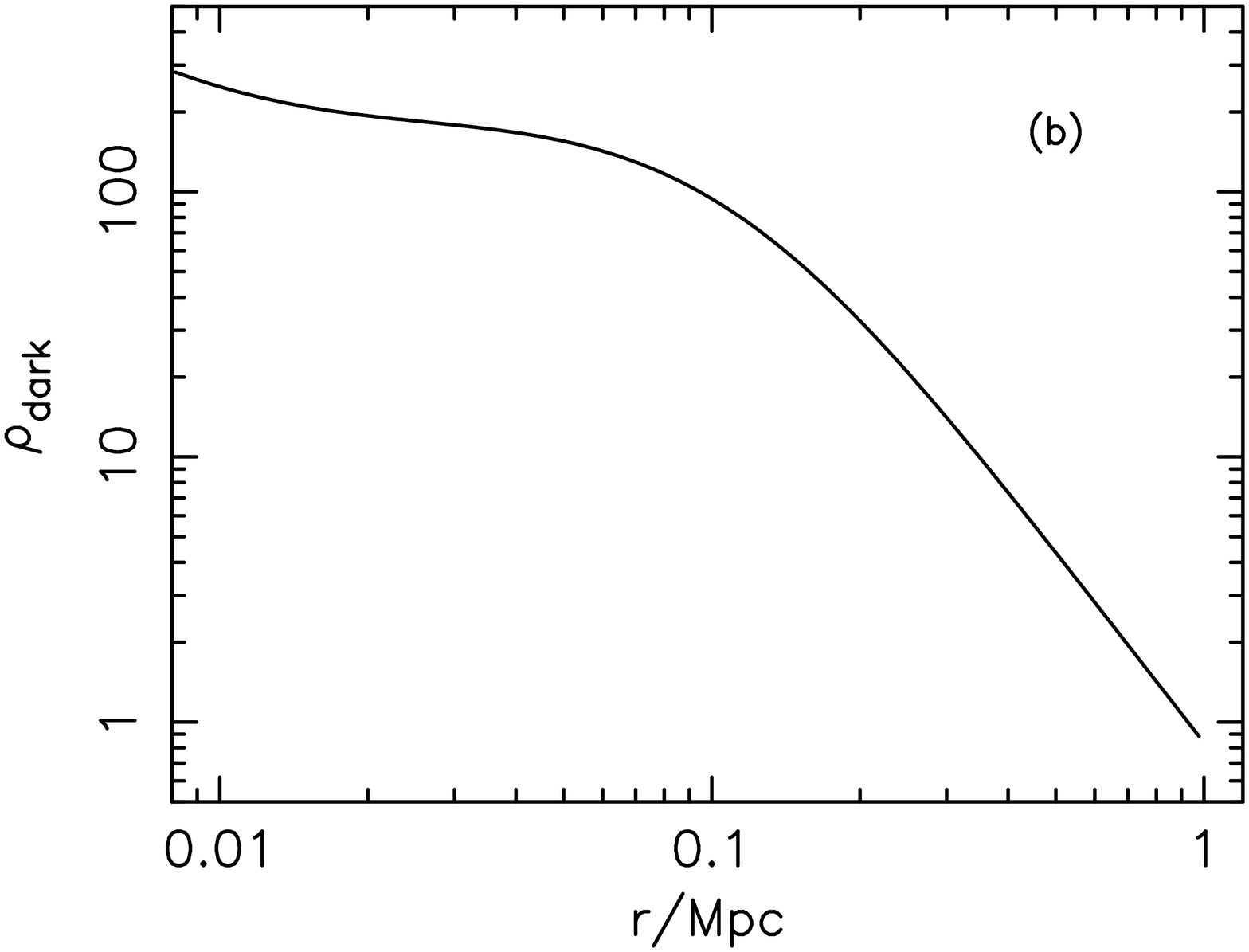,width=7.2cm,angle=270}
}
\caption{The $k=1$ cooling flow solution for A85; (a) $\Sigma$ (dashed
line) and $\tau$ (solid line), (b) the density of the gravitating matter.}
\label{fig:a85k001}
\end{figure}

I will now give an example of the application of the theory to the
cooling flow in the cluster A85.  The emissivity profile obtained with
the ROSAT HRI (after correction for absorption and the instrumental
energy response) were kindly supplied to me by Clovis Peres.  It is
shown in 12 arcsec bins in \Fig~\ref{fig:a85xi} together
with a simple broken power-law fit,
\be
\xi\propto\left[\left(r\over0.12\,{\rm Mpc}\right)^{1.18}+
\left(r\over0.12\,{\rm Mpc}\right)^{2.83}\right]^{-1}.
\ee
The asymptotic slope of $\xi$ as $k\mapsto0$ is very close to the minimum
permitted for $k=1$ which suggests that the solution will require an
inner core in the mass distribution.  This is illustrated in
\Fig~\ref{fig:a85k001}.  Note that $\Sigma$
drops very close to zero at $r=10\,$kpc (it tends to a small constant
value within this radius).  The gravitational density profile (\ie
that of the total mass, not just the gas) is well-fit at radii greater
than 20\,kpc by a King model,
\be
\rho_{\rm grav}
\propto\left[1+\left(r\over0.11\,\mbox{Mpc}\right)^2\right]^{-1.25}.
\ee
Within this radius the density is poorly constrained.  Although it
appears to rise abruptly, only a small change in the slope of $\xi$
would cause it to level off or even fall---all we know for sure is
that the virial temperature becomes very small.  Note also that there
is only one bin within 20\,kpc and this one is most likely to be
affected by smoothing by the point-spread function, uncertain correction
for excess absorption, etc.

The temperature of the gas is approximately constant outside the core
radius, but drops by a factor of five in to 10\,kpc.  A temperature
decline in the centre of clusters is typical of cooling flows observed
by ASCA.

Note that the slope of the mass-deposition profile, $\tau$, is close
to unity within the cooling radius, $r_{\rm cool}\approx150\,$kpc.
This radius is not a special one for our solutions as we have assumed
the cooling flow solution holds everywhere.  For this reason the
asymptotic slope of the gravitational density profile at large radii
should be taken with a pinch of salt.

\begin{figure}
\parbox{7.2truecm}{
\psfig{figure=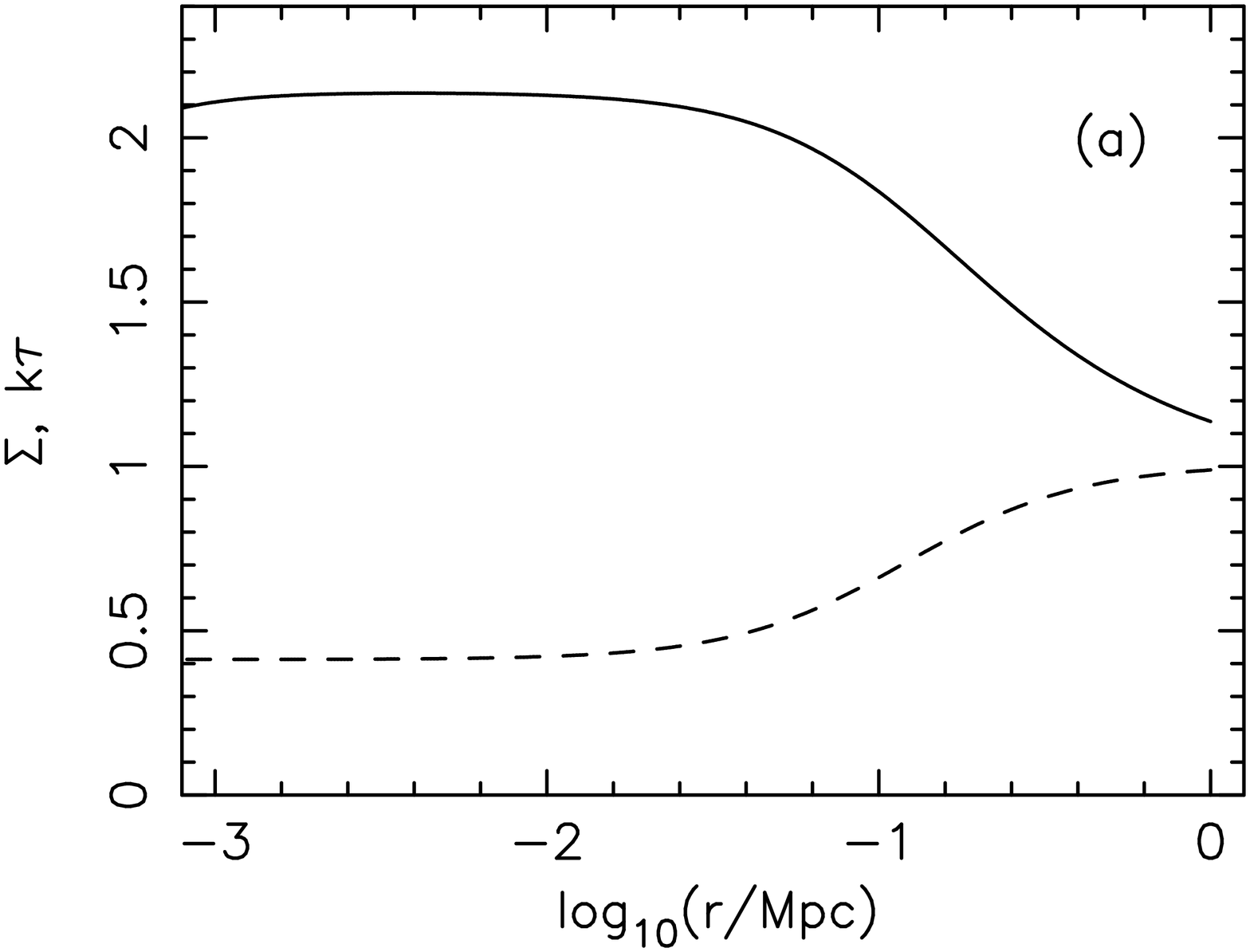,width=7.2cm,angle=270}
}\parbox{7.2truecm}{
\psfig{figure=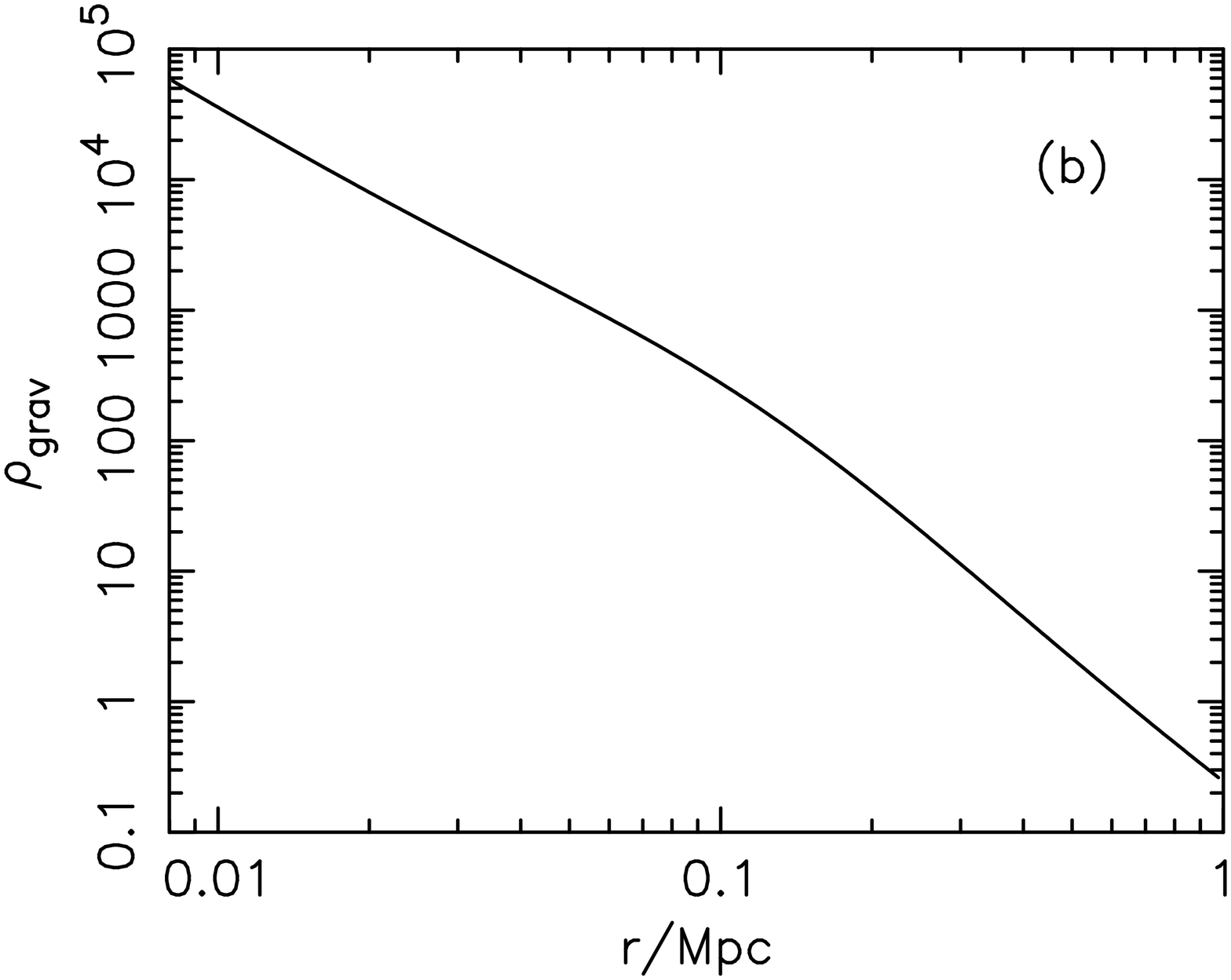,width=7.2cm,angle=270}
}
\caption{The $k=\infty$ cooling flow solution for A85; (a) $\Sigma$ (dashed
line) and $\tau$ (solid line), (b) the density of the gravitating matter.}
\label{fig:a85k100}
\end{figure}

The corresponding solution for $k=\infty$ is shown in \Fig~\ref{fig:a85k100}.
There in no core in the gravitational mass profile in this
case, with $\rho_{\rm dark}$ rising as $r^{-2}$ all the way into
10\,kpc.  This is reflected in the temperature profile, however, which
also rises by a factor of 5 between 150 and 10\,kpc.

I conclude that A85 has a maximum core radius of 120\,kpc and that
self-similar solutions with low values of $k$ provide more plausible
temperature gradients than those with high values.  If the mass
profile has a constant-density core then $k=1$, as predicted above.

\section{Conclusions}

\begin{itemize}
\item Cooling flows are known to be multiphase and the icm of clusters
may well be multiphase throughout.  This can affect mass measurements.
\item I have rederived the cooling flow equations for self-similar
density distributions.  Two of these in particular are expected to
bound the behaviour of all possible flows.
\item The steady-state cooling flow equations are \emph{not}
compatible will all conceivable emissivity profiles.  Thus they can be
used as a test of the theory.
\item The solutions provide bounds on $M(r)$ within the cooling radius
and given $k$ can be used to measure $M(r)$.
\end{itemize}

It would be very useful to apply the cooling flow models discussed
here to galaxies and clusters with well-resolved cooling flows and
especially those with measured mass-profiles or temperature gradients.

\acknowledgments

I would like to thank Clovis Peres for supplying me with the data for
A85.  This work was carried out while I was a Nuffield Foundation
Science Research Fellow.


\begin{references}
\reference Gunn K. F., Thomas P. A., 1996, \mnras, in press
\reference Nulsen P. E. J., 1986, \mnras, 221, 377
\reference Thomas P. A., 1988a, in {\bf NATO ASI} Cooling flows in
   clusters and galaxies, A. C. Fabian, Dordrecht: Kluwer, 361
\reference Thomas P. A., 1988b, \mnras, 235, 315
\reference Thomas P. A., Fabian A. C., Nulsen P. E. J. N., 1987,
   \mnras, 228, 973
\reference Walker T. P., Steigman G., Schramm D. N., Olive K. A.,
   Kang H. -S., 1991, \apj, 376, 51
\reference White D. A., Fabian A. C., 1995, \mnras, 273, 72
\reference White S. D. M., Navarro J. F., Evrard A. E., Frenk C. S.,
   1993, Nature, 366, 429
\end{references}
\end{document}